\documentstyle[aps]{revtex}

\begin{document}
\title{Inclusive Higgs boson and dijet production via Double Pomeron exchange}
\author{M. Boonekamp\thanks{%
CEA, SPP, DAPNIA, CE-Saclay, F-91191 
Gif-sur-Yvette Cedex,
France}, R. Peschanski\thanks{%
{\it idem,}  SPhT} and C. Royon\thanks{%
{\it idem,} SPP and  BNL, Upton, NY (USA),  Texas U at Arlington (USA) }}
\maketitle

\begin{abstract}
We evaluate Higgs boson and dijet cross-sections at the Tevatron collider via 
Double 
Pomeron exchange when accompanying particles  in the central region are taken 
into account. Such {\it inclusive} processes, normalized to the observed dijet 
rate observed at run I, noticeably increase the predictions for tagged 
(anti)protons in the run II with respect to {\it exclusive} ones, with the 
potentiality of Higgs boson detection.  
\end{abstract}
\bigskip
{\bf 1.}
Not only recently, Higgs boson and dijet production via Double Pomeron (DP) 
exchange 
have attracted 
attention. It has been shown \cite{bi90} that a non negligeable 
fraction of events with double colour singlet  exchange 
features may 
appear. The calculation was based on a Pomeron model formulated as  a pair of 
non-perturbative QCD gluons. However, the lack of solid QCD theoretical 
framework for diffraction makes a theoretical determination difficult. 
Predictions with many different mechanisms appeared \cite{all} since then  
without clear consensus on the expected rate at hadron colliders. An 
evaluation of experimental possibilities \cite{al00} has given  
the prediction of a discovery potential  for the Higgs boson through diffraction 
using 
outgoing (anti)proton tagging and a missing mass method. However, this 
evaluation again  
strongly depends  on the theoretical framework.

In the present paper our aim is to give predictions based on {\it inclusive} 
Higgs boson and dijet production at the Tevatron collider via DP
exchange. Indeed, so far, the predictions have been based on {\it exclusive} 
production without taking into account the accompanying particles in the central 
rapidity region. This accompanying radiation is  unavoidable and is present  in 
dijet production,  
a DP process already detected at the Tevatron \cite {CDF0}. This is obvious from 
the observed dijet over total mass fraction spectrum 
which is very 
different from the one expected in absence of radiation. 

In fact, the evaluations  contained in the 
present paper may have a triple 
phenomenological interest. {\bf i)} Taking into account the accompanying 
particles  
allows one to normalize the theoretical predictions to the observed dijet rate 
and thus can be translated into a more constrained prediction for the Higgs 
boson. 
{\bf ii)} The {\it inclusive} cross-sections are expected to be larger than the  
{\it exclusive} production mode {\bf iii)} The
signal/background ratio  has to be evaluated differently  in {\it inclusive} and 
{\it 
exclusive} events. 

In the following we use the Bialas-Landshoff {\it exclusive} model for Higgs 
boson and 
heavy flavor jet production \cite{bi90} 
as a starting point. We  modify it in order to 
take into account {\it inclusive} Higgs boson and dijet production and the 
addition of 
 light 
quark jets and gluon jets (not 
included in the analysis of Ref. \cite{bi90}) . We show that we are 
able 
to reproduce up to a scale (which will be fixed from experiment) the observed 
distributions, in particular  the dijet mass fraction spectrum. Using the same 
framework, we give predictions for 
{\it inclusive} Higgs boson  production via  DP exchange  at the 
Tevatron. 

The main lesson of our study is that, if the signal/background ratio can be 
maintained at a 
reasonable level when associated central radiation is allowed, an 
interesting discovery potential for the Higgs boson particle at the Tevatron run 
II 
can be expected in double (anti)proton tagged experiments. It is materialized in 
table I for the number of events as a function of $M_H$ depending on 
experimental cuts and decay products.

 Note that theses estimates are sizeably higher than   the more recent {\it 
exclusive} production predictions \cite {all}. The inclusion of all {\it 
inclusive } production modes of dijets  allow us to fix 
the estimated proportion of quarks over gluon jets. It is  small but not 
negligeable, contrary to  the {\it exclusive } production. Thus DP production of 
dijets 
cannot be called   a pure ``gluon factory'' (cf. last paper of 
Ref. \cite {all}).
One reason of this difference  is 
for {\it exclusive} 
production the helicity 
conserving suppression of dijet production into quark jets vs. gluon jets which 
is absent in {\it inclusive} 
production. This re-normalization, in turn, enhances the prediction for Higgs 
boson 
production via the $\bar t t$ loop.

\bigskip
{\bf 2.} 
Let us introduce the formulae for Higgs boson and dijet production 
cross-sections  via 
 DP exchange. In the {\it exclusive} channel \cite{bi90} one writes;
\begin{eqnarray}
& &
d\sigma_H^{excl} = C_H\left(\frac s {M_H^2}\right)^{2\epsilon} \ \delta 
\left((1\!-\!x_1)(1\!-\!x_2)\!-\!{M^2}/s \right) \ 
\prod _{i=1,2} \ \left\{\frac {dx_i}{x_i}\ d^2v_i\ 
(1\!-\!x_i)^{\alpha' v_i^2}\ \exp \left(-2\lambda v_i^2\right)\right\}\ , 
  \nonumber \\
& &
d\sigma_{\bar Q Q}^{excl} = C_{\bar Q Q} \left(\frac s {M_{\bar Q 
Q}^2}\right)^{2\epsilon} F_{\bar Q Q}(\rho)\ \delta ^2 
\left(\sum_{i=1,2}(v_i\!+\!k_i)\right) 
\prod _{i=1,2} \left\{\frac {dx_i}{x_i}\  d^2v_i\ d^2k_i\ 
(1\!-\!x_i)^{\alpha' v_i^2}\ \exp \left(-2\lambda v_i^2\right)\right\}
 \ ,
 \label{dsigma}
\end{eqnarray}
for a  Higgs boson of mass $M_H$ and  two heavy quark jets (of total mass 
$M_{\bar Q 
Q}$), respectively. $\alpha (t)=1+\epsilon+\alpha't$ for the Pomeron trajectory 
($\epsilon\sim 
.08,\alpha'\sim .25 GeV^{-2}$), $x_{1,2}\ (>.9)$ are the fraction of momentum 
of 
the outgoing $p$ and $\bar p$, $v_{1,2},$ their 2-transverse momenta, 
$k_{1,2},$ those of the outgoing quark jets, $\lambda \sim 4 GeV^{-2}$ the slope 
of the Pomeron $p \bar p$ coupling, and the constants $C_H , C_{\bar Q Q}$ are 
 normalizations containing various factors related to the hard matrix elements 
together with a common non-pertubative factor 
  $G^8$ due to
 the  non-pertubative gluon coupling \cite{bi90}. Thus, the ratio $C_H / 
C_{\bar Q Q}$ is well defined while the overall 
normalization is not known, and will be determined by direct comparison with 
data.

For  heavy quarks of mass $m_Q$ and transverse mass 
$m_{T1,2}$  the hard contribution reads
\begin{equation}
F_{\bar Q Q}(\rho) = \frac 
{\rho\ (1-\rho)}
{m_{T1}^2m_{T2}^2}\ ,\ \rho\equiv \frac {4m_Q^2}{M_{\bar Q Q}^2}
\label{rho}
\end{equation}
which, up to colour factors included in the $C_{\bar Q Q},$ stands for  
the $gg\to 
\bar Q Q$ hard cross-section in the model. It is worthwhile to note that 
this cross 
section is proportional to $m_Q^2,$
and thus is quasi zero for light quarks. This reflects the known  zero helicity 
constraint  of the quark jet production mechanism of Ref. \cite{bi90} (For light 
quarks a zero appears in the forward anti(proton) direction, see dijet 
papers in \cite {all}).

The method we propose to evaluate the {inclusive} production mechanisms for 
Higgs boson and dijets (including light quarks and gluons) is simple and not 
restricted to the Bialas-Landshoff model, even if we use it as a starting point. 
The main driving idea is   that whatever the mechanism 
and normalization  of 
the diffractive part of the process could be, the ``hard'' partonic interaction 
in the central region depends on the probability of finding the initial partons 
(quarks and gluons) in the central region at the short time. Since the overall 
partonic configuration is produced initially by the long-range, ``soft'' DP 
interaction, 
we will assume that, up to a normalization, 
the {\it inclusive} cross-section is the convolution of the  ``hard'' $partons 
\to Higgs \ boson, partons \to jets$  subprocesses by the probability 
of finding these partons in both Pomerons, see Fig.1. The working hypothesis is 
that the expected factorization 
breaking  from soft color interaction will essentially affect  
the normalization through a renormalization of the Pomeron fluxes, which are not 
the same as in hard diffraction at HERA.

In this framework, QCD radiative corrections is the natural origin  of {\it 
inclusive} production \cite{pe01}. Indeed, our 
ansatz remarkably reproduces the dijet mass fraction seen in 
experiment which is obviously different from {\it exclusive} production, see 
Fig.2. 
 
 Hence,  the Higgs boson and dijet {\it inclusive} cross-sections become:
\begin{eqnarray}
d\sigma_H^{incl} &=& G_P(x^g_1,\mu)G_P(x^g_2,\mu)\ \frac {dx^g_1}{x^g_1}\ \frac 
{dx^g_2}{x^g_2}\ d\sigma_H^{excl} (s \to x^g_1x^g_2\  s), 
  \nonumber \\
d\sigma_{JJ}^{incl} &=& C_{\bar Q Q}\left(\frac {x^g_1x^g_2  s }{M_{\bar J 
J}^2}\right)^{2\epsilon}\!\! F_{JJ}\ \delta ^2 
\left(\sum _{i=1,2} (v_i\!+\!k_i)\right) \prod _{i=1,2} \left\{G_P(x^g_i,\mu)\ 
\frac 
{dx^g_i}{x^g_i}\ \frac {dx_i}{x_i} d^2v_id^2k_i\ (1\!-\!x_i)^{\alpha' v_i^2}\ 
\exp 
\left(-2\lambda v_i^2\right)\right\}
\ , 
 \label{dinclu}
\end{eqnarray}
where $x^g_1, x^g_2$ define the fraction of the Pomerons' momentum carried by 
the 
gluons involved in the hard process, see Fig.1, and $G_P(x^g_{1,2},\mu),$ are, 
up to a normalization, the gluon 
structure function in the Pomerons extracted  from HERA experiments, see  
\cite{ba00}.  
$\mu^2$ is the hard scale (for simplicity kept fixed at 
$75 GeV^2,$ the highest value studied at HERA). We neglected the 
processes 
initiated by quarks in the Pomeron which are negligeable in \cite{ba00}.

The dijet  cross-section  $\sigma_{JJ}$  depends on 
two``hard'' cross-sections $gg\to \bar 
Q^{(i)} Q^{(i)}$ and $gg\to gg$  \cite{co83}. This gives for  5 quark flavors 
$(i= 1 \cdot 5)$
\begin{eqnarray}
 & &
F_{JJ}= \sum _{i} F_{\bar Q^{(i)}Q^{(i)}} \left(\rho^{(i)}\right) + {27} \ 
F_{gg}\left(\rho^{gg}\right)\ ; 
\ \rho^{(i)}\equiv \ \frac {4\ m_{T1}^{(i)}m_{T2}^{(i)}}{M_{\bar Q^{(i)} 
Q^{(i)}}^2}  ;\ 
\rho^{gg}\equiv \ \frac {4\ p_{T1}p_{T2}}{M_{gg}^2}\ ,\nonumber\\
 & &
F_{\bar Q^{(i)}Q^{(i)}} \equiv \frac { \rho^{(i)}} {m_{T1}^{(i)}m_{T2}^{(i)}}\ 
\left(1-\frac 
{\rho^{(i)}}2\right)\left(1-\frac 
{9}{16}\ \rho^{(i)}\right)
 \ ;\ F_{gg} \equiv \ \frac 1 {p_{T1}^2p_{T2}^2}\ 
\left(1-\frac {\rho^{gg}}4\right)^3\ .
\label{rhoi}
\end{eqnarray}
 ${27}$ stands for the  colour factor 
fraction between gluon jets and quark jets partonic cross-sections \cite{co83}.
 Note that from now on , the $gg\to \bar Q^{(i)} Q^{(i)}$ cross section is 
proportional to {\it transverse} and not to rest quark masses. Thus, all 5 quark 
flavors sizeably contribute to the dijet cross-section. The quark  contribution 
remains 
smaller, due to the color factor and to the   $\sim 1 /{p_T^4}$  behaviour of 
the 
gluon exchange cross-section compared to the  $\sim 1/(p_T^2 M^2_{\bar Q^{(i)} 
Q^{(i)}})$ one of  the quark exchange cross-section but is no more suppressed by 
the helicity constraint. 

{\bf 3.} The first step we perform is a comparison of our results with the
measurements performed at Tevatron, in the CDF experiment \cite{CDF0}.
To this end, we interfaced our generator with a fast simulation of the
D0 and CDF detectors, namely {\tt SHW} \cite{shw}. We chose as gluon
content of the pomeron the result of the H1 ``fit 1'' performed in
Ref. \cite{ba00}. Note that the gluon density is  used only up to 
 the normalisation of the flux. We  normalize our dijet
cross section prediction with the CDF one. 

We first compare our results for the dijet mass fraction with the
measurement of the CDF collaboration \cite{CDF0} in DP
events.
A tagged antiproton with $0.035
\le \xi_{\bar{p}} \le 0.095$ and $|t| < 1$ GeV$^2$ was required where $\xi_{\bar 
p,p} \equiv 1\!-\!x_{1,2}$
is the momentum faction of the $\bar p,p$ carried by the corresponding pomeron.  
This quantity
is reconstructed using the roman pot detectors installed by the CDF
collaboration. After the CDF cuts to tag an antiproton and the fast simulation 
of
the detector, we obtain a cross section of 14.4 $nb$, to be compared with the 
CDF
measurement of 43.6 $nb$.
At the present stage, at least two aspects are lacking in our study :
firstly, our cross-section formulae do not include dissociation of the
outgoing (anti)protons. Secondly, some (difficult to evaluate) fraction
of the proton-tagged events seen in experiment are in fact fake tags,
coming namely from excited N* states leaving an energetic decay proton
in the detector. Both effects contribute to enhance the measured
cross-section, and thus the difference between our prediction and the
measured value
\footnote{Note that the CDF cross section
was measured after kinematical cuts on the proton side: 
$0.01 \le \xi_p \le 0.03$. This cut is difficult to reproduce using a
fast simulation of the detector since it is very sensitive to the energy
measurement inside the main CDF detector (the proton is not tagged). Thus, we 
did
not apply this cut on our cross section. If one tries to apply these cuts
in our fast simulation program, one obtains a cross-section two or three times
smaller depending on the energy corrections. We chose not to apply these
additional factors to be on the safe side concerning our predictions
on  the numbers of Higgs boson events. 
Thus, our Higgs boson event rates may be conservative by a
factor 2 to 3 if one takes into account this effect.}.
We thus scale up our cross-sections by a factor $43.6/14.4 \sim 3$. As 
shown in
Fig. 2, the dijet mass fraction spectrum is well reproduced. The CDF
measurement could clearly not be described without radiation since
the obtained dijet mass fraction would peak near one, up to detector resolution 
effects. 

We can now give predictions for the Higgs boson production cross sections in 
double
diffractive events, by scaling our results by the abovedetermined factor . The 
results are given in Table 1, first column. We note the high values 
of
the cross-sections, which predict\footnote{The expected luminosity
is between 20 and 25 $fb^{-1}$ per experiment for run II.} more than 10 events 
per $fb^{-1}$ for
a Higgs boson mass below 140 GeV.

After interfacing the generator with the fast simulation package 
{\tt SHW} \cite{shw}, we can estimate the rates which could be observed 
in the experiments. The experimental resolution and acceptances of the
roman pot detectors have been chosen to be similar to the D0 ones for
dipole detectors, namely the $t$-resolution is $0.1 \sqrt{t}$,
$t$-acceptance $|t| \le$ 0.5 GeV$^2$, 
$\xi$-resolution 0.2 \%, and the $\xi$-acceptance 100\% if $\xi >0.04$,
0\% if $\xi < 0.01$ and linear between 0 and 100\% if $0.01 < \xi < 0.04$ 
\cite{d0fpd}. The tagging efficiency (see column 2 of Table 1) is quite
good if one uses dipole detector on each side. To be able to trigger these
events, some activity inside the central detector will be required, and
we give in Table 1, third column, the number of events after requiring at least
two jets of $p_T > 30$ GeV \footnote{This allows us to enhance our 
signal/background
ratio (the $b \bar{b}$ diffractive background without any cuts is estimated to
be about 2.3 $10^7$ events per $fb^{-1}$, and becomes about 8.3 $10^4$ events 
after 
those
$p_T$ cuts), as shown in Fig.3 c,d. When we compare Fig.2 c,d, we also note that 
radiation is more important for dijet events than for
Higgs boson events, since the distribution for dijet events is more shifted to
the left after radiation than the Higgs boson one. Hence, the $p_T$
cut is more efficient after radiation.}. 

To enhance the signal to background ratio, it is possible to cut on the proton 
and antiproton
tagged energy at 930 GeV (see Fig.3a), on the jet topologies (the jets coming
from Higgs boson events are more central) and
on the reconstructed mass
distribution using the missing mass method \cite{al00}  
(see Fig.3b, and
Fig.4). We slightly modified the original method to partly take into
account radiation and define 
$M_H = \sqrt{\xi_p \xi_{\bar{p}} S} \frac{E_{jet1} + E_{jet2} + E_p
+E_{\bar{p}}}{2 E_{beam}}$ where the $E_{jeti}$ are the leading two jets
energies, $E_p$ and $E_{\bar{p}}$ the tagged $p$ and $\bar{p}$ energies. 
We notice that the missing mass method is not working so nicely
when radiation is included. It is however still a competitive method to reduce 
background and reconstruct
the Higgs boson mass\footnote{Due to radiation effects which escape mostly into
the beam pipe, the missing mass method cannot work as it stands and
is modified by radiation. However, it will be very useful to perform
constrained fits and to have different ways 
of reconstructing the Higgs boson mass.}. The background 
over signal ratio will be detailed in Ref \cite{bo01}.
Since we obtain quite high cross-sections, other Higgs boson decay channels
with smaller branching ratios, like $H \rightarrow \tau^+ \tau^-$ (about 10\%
of Higgs boson decay, see table 1) or 
$H \rightarrow W^+ W^-$ are of very high interest since the expected
background is very small. This will be further discussed  in Ref. \cite{bo01}.

\bigskip \bigskip \bigskip

{\bf ACKNOWLEDGMENTS}

\bigskip\bigskip

 We acknowledge fruitful discussions with M. Albrow, A. Bialas, A. 
Brandt, A. De 
Roeck, V. Khoze, L. Schoeffel.
 
\vspace{1cm}
{\bf TABLE} \newline
\begin{table}[b]
\begin{center}
\begin{tabular}{|c||c|c|c|c|c|} \hline 
$M_{Higgs boson}$ &(1)&(2)&(3)&(4)&(5) \\
\hline\hline
100 & 26.6  & 18.5  & 5.7  & 1.9  & 0.2 \\
110 & 21.6  & 14.0  & 5.3  & 1.3  & 0.7 \\
120 & 17.4  & 9.8   & 4.8  & 1.0  & 1.9 \\
130 & 13.8  & 6.1   & 3.2  & 0.6  & 3.3 \\
140 & 10.6  & 2.9   & 1.8  & 0.3  & 4.2 \\
150 & 8.0   & 1.0   & 0.8  & 0.1  & 5.0 \\
160 & 5.7   & 0.2   & 0.1  & 0.0  & 4.5 \\
170 & 3.7   & 0.0   & 0.0  & 0.0  & 2.9 \\
\hline
\end{tabular}
\bigskip 
\caption{{\it Number of Higgs boson events for 1 $fb^{-1}$.} The first column 
gives
the
number of events at the generator level (all decay channels included), and 
the other columns after
a fast simulation of the detector. The second colum gives the number of
events decaying into $b \bar{b}$
tagged in the dipole roman pot detectors (see text), the third one
requiring
additionally at least two jets of $p_T > 30 GeV$,
the fourth one gives the
number of reconstructed and tagged events when the Higgs boson decays into
$\tau$, and the fifth one when the Higgs boson decays into $W^+ W^-$
(in this channel, the background is found to be negligible).}
\end{center}
\end{table}

{\bf FIGURES}

\bigskip  

\input epsf \vsize=8.truecm \hsize=10.truecm \epsfxsize=8.cm{%
\centerline{\epsfbox{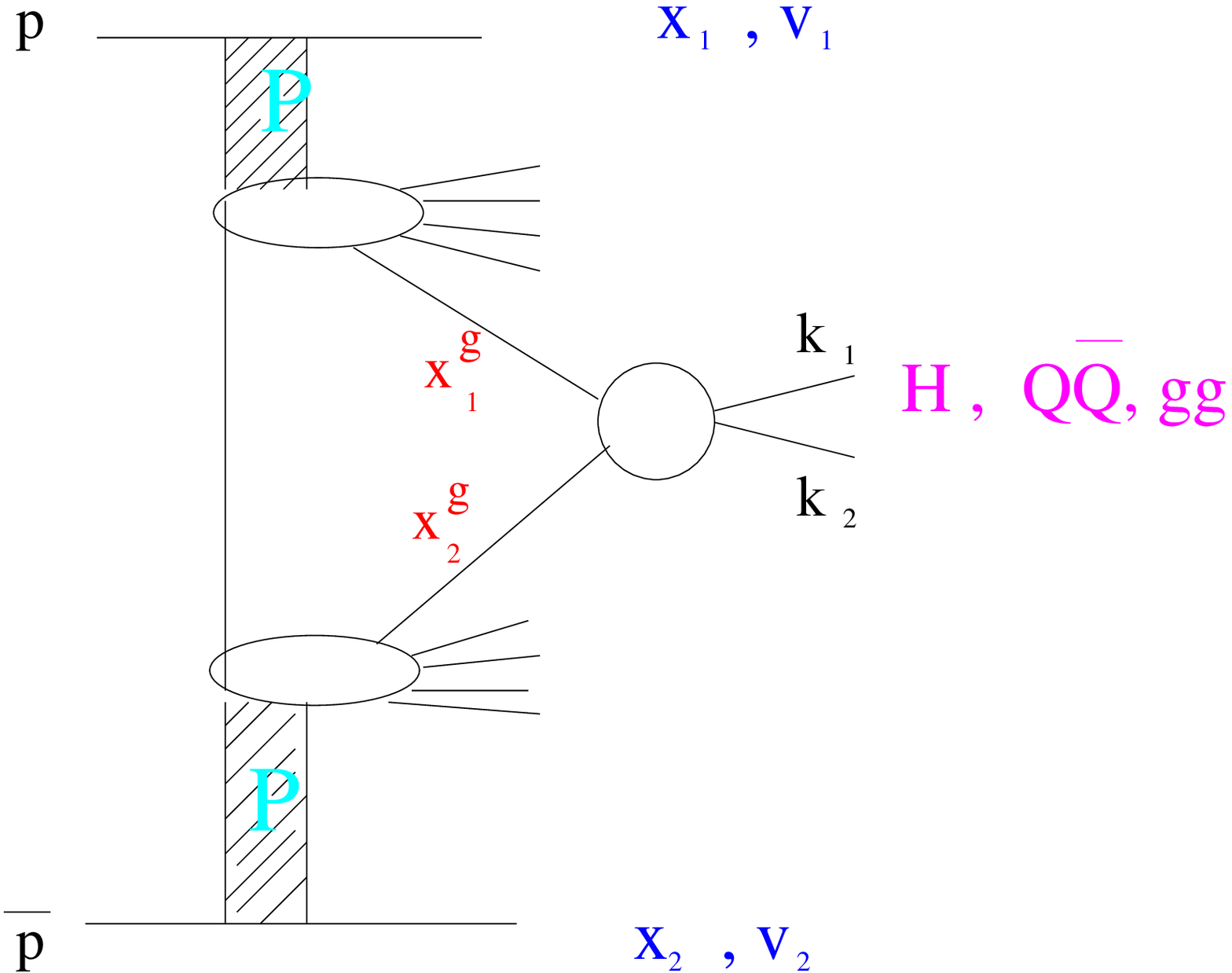}}}\bigskip 
 {\bf Figure 1} 

{\it Inclusive production schemme.}

$x_i,v_i$ are the longitudinal and transverse 2-momenta of the diffracted 
(anti)proton, $x_i^g,$ the Pomeron  fraction momentum brought by the gluons 
participating in the hard cross-section, $k_i,$ the transverse 2-momenta of the 
outgoing jets in the central region from quarks, gluons or the $\bar b b$ decay 
products of 
the Higgs boson.

\input epsf \vsize=8.truecm \hsize=10.truecm \epsfxsize=8.cm{%
\centerline{\epsfbox{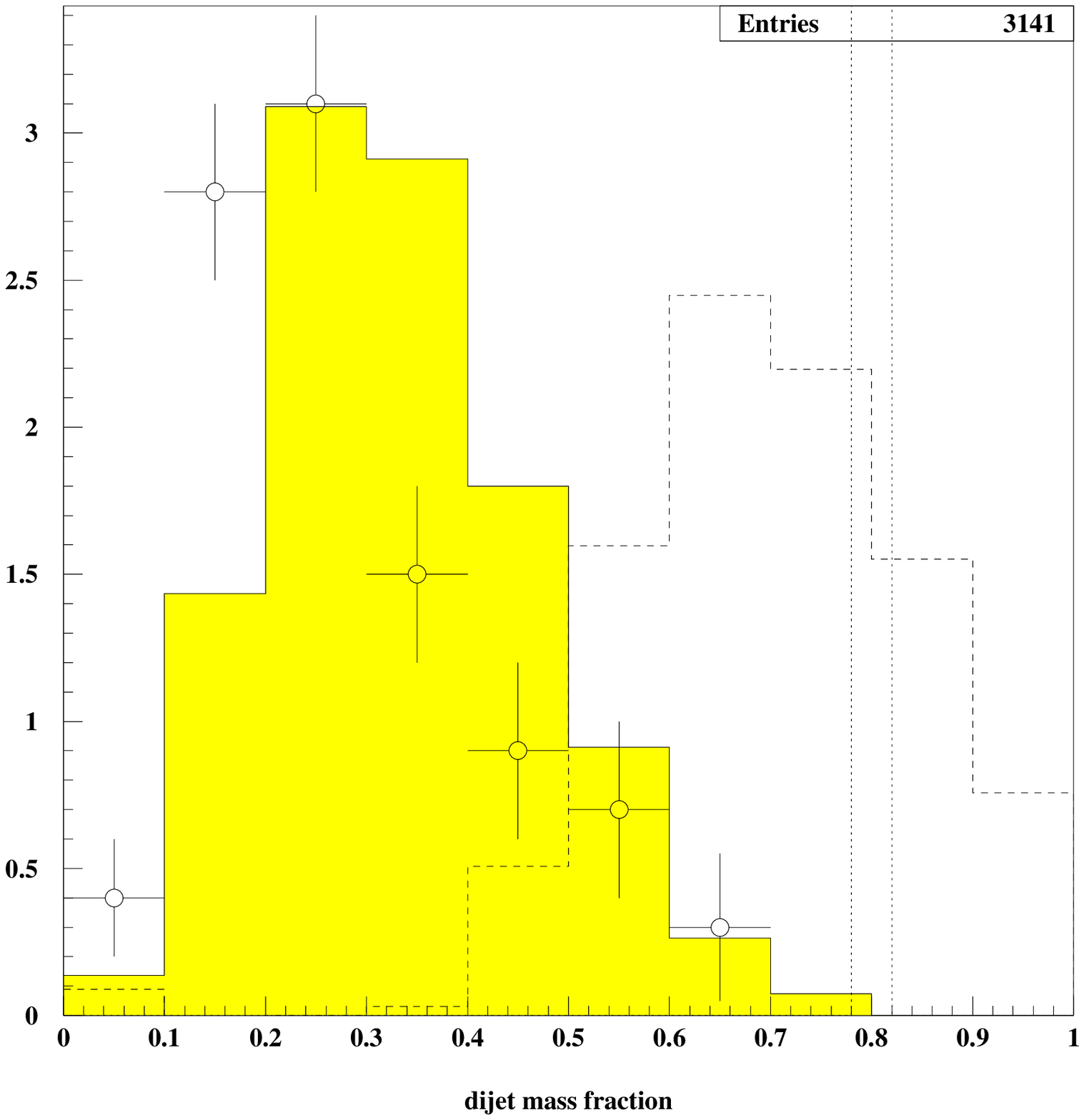}}}\bigskip 
 {\bf Figure 2} 

{\it Comparison of the dijet mass fraction obtained in our model and CDF data
(open circles).}

\noindent {\it With radiation:}  The shaded distribution is the   dijet 
distribution  after
radiation and simulation of the detector. \newline {\it Without radiation:} 
Dotted line: distribution at generator level; Dashed line: after simulation of 
the detector.

\input epsf \vsize=8.truecm \hsize=10.truecm \epsfxsize=8.cm{%
\centerline{\epsfbox{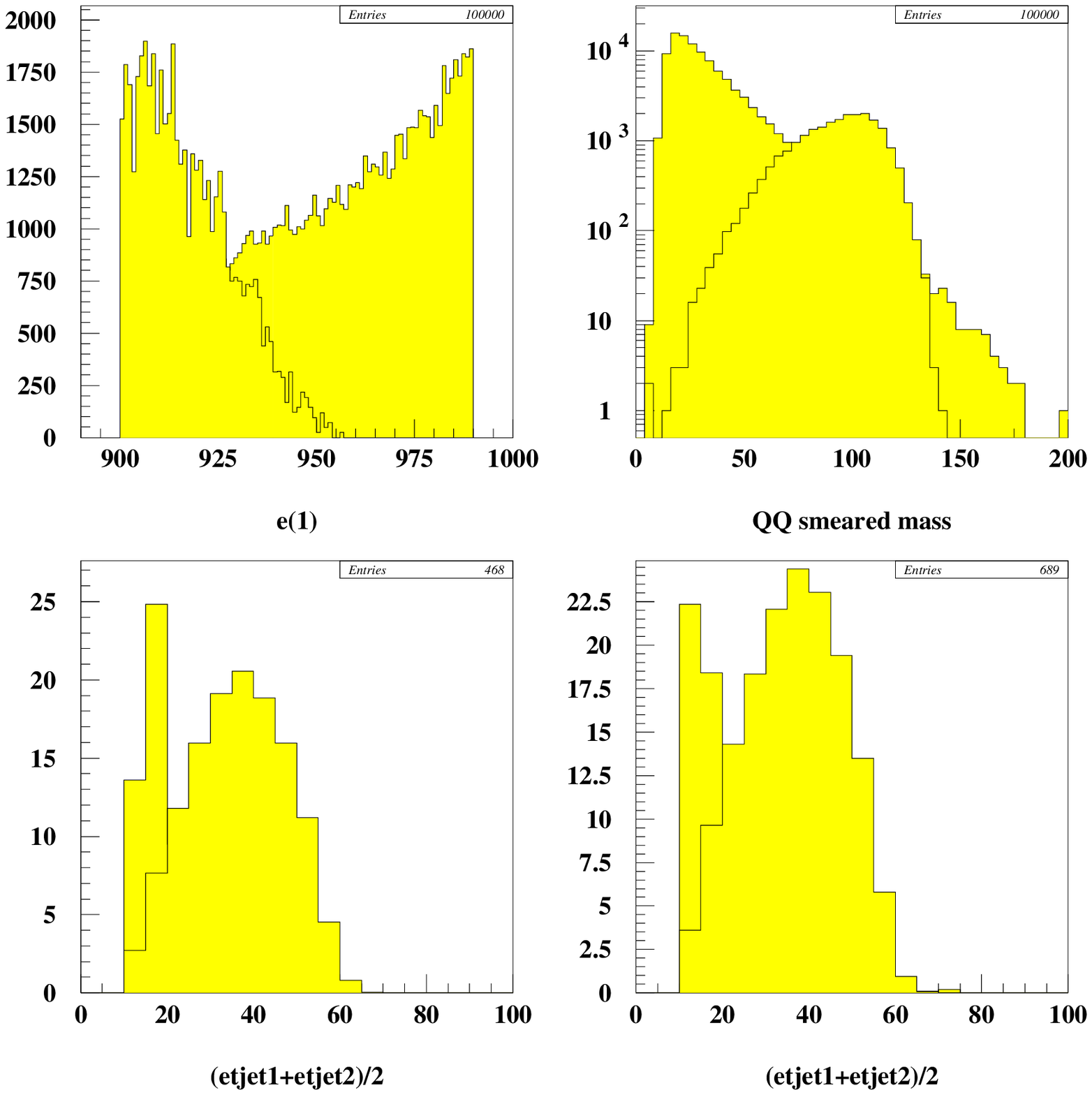}}}\bigskip 
 {\bf Figure 3} 

{\it Comparison of diffractive dijet and Higgs boson distributions.}

the Higgs boson distributions are here always on the right of the  $\bar b b$ 
distribution.
Fig (3a, up left) gives the energy distribution of the tagged proton or 
antiproton,
Fig (3b, up right) the leading dijet mass distribution, Fig (3c, down left) the 
mean leading two jet transverse energy in case radiation is added, and Fig
(3d, down right)
is the same plot without radiation. Normalisation is arbitrary in all
figures.

\input epsf \vsize=8.truecm \hsize=10.truecm \epsfxsize=8.cm{%
\centerline{\epsfbox{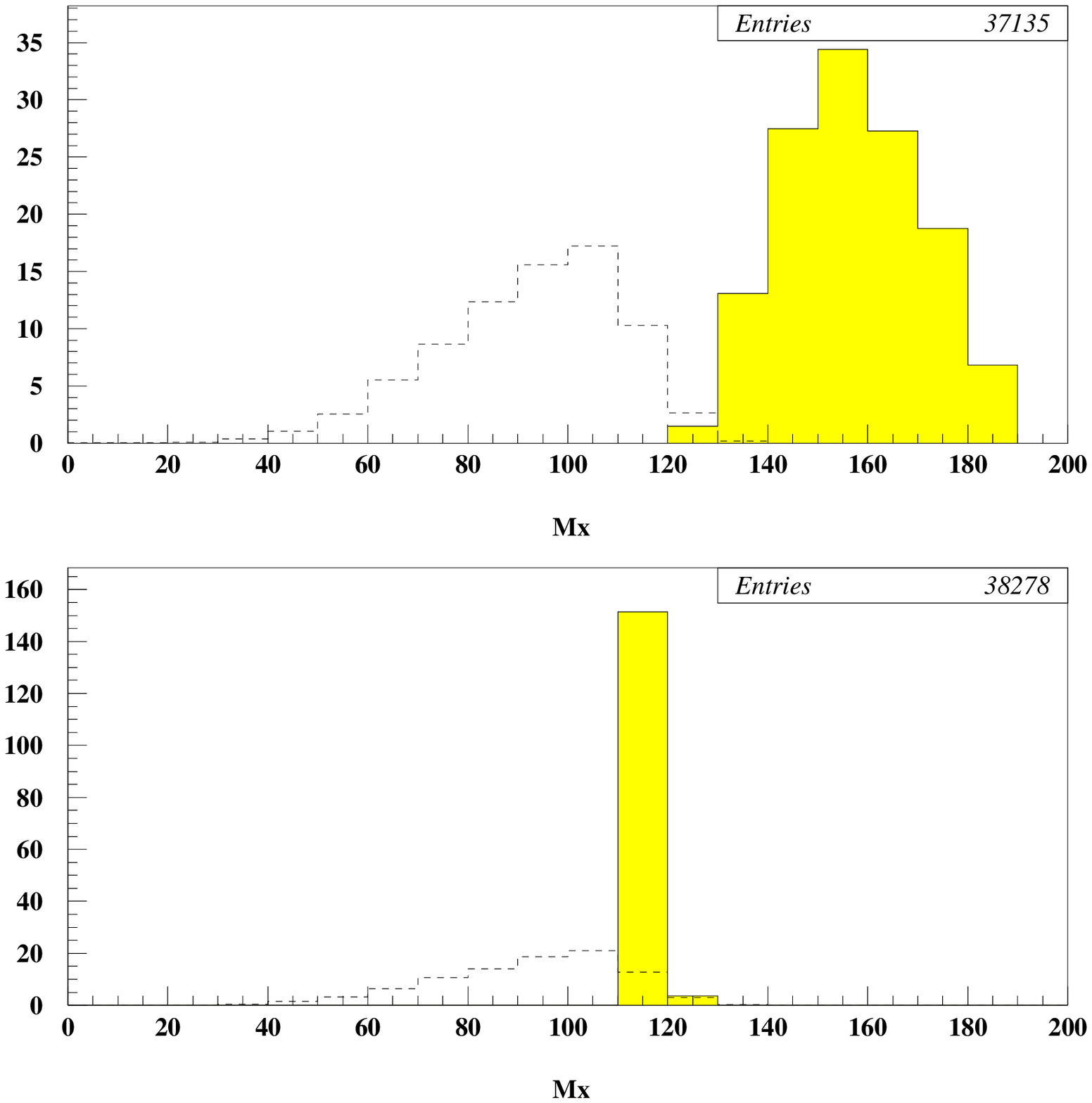}}}\bigskip 
 {\bf Figure 4} 

{\it Higgs boson mass reconstruction.}
On top, with radiation, on bottom, without. 
The grey distribution is the result of the missing mass method, and the
dashed line is the leading dijet mass distribution.

\end{document}